\documentclass[final,5p,times,numbers,compress]{elsarticle}

\usepackage{amssymb}
\usepackage{hyperref}
\usepackage{lipsum}

\usepackage{amsmath}

\journal{Physics Letters B}

\begin{document}

\begin{frontmatter}



\title{Constraints on lepton-flavor mixing with third-generation new physics}


\author[first]{Sebastiano Covone}
\ead{sebastiano.covone@physik.uzh.ch}
\affiliation[first]{organization={Physik-Institut, Universitat Zürich},
            addressline={Winterthurerstrasse 190}, 
            city={Zürich},
            postcode={CH-8057}, 
            country={Switzerland}}

\author[second,secondbis]{Pol Morell}
\ead{pmorell@icc.ub.edu}
\affiliation[second]{organization={Departament de Física Quàntica i Astrofísica},
            addressline={Martí Franquès 1}, 
            city={Barcelona},
            postcode={E08028}, 
            country={Spain}}
\affiliation[secondbis]{organization={Universitat de Barcelona},
            addressline={Martí Franquès 1}, 
            city={Barcelona},
            postcode={E08028}, 
            country={Spain}}

\author[first]{Arianna Tinari}
\ead{arianna.tinari@uzh.ch}

\begin{abstract}
We study the implications of an approximate $U(2)^5$ flavor symmetry at the TeV scale, under the assumption of new physics predominantly coupled to the third-generation fermions, focusing on the breaking of the $U(2)_\ell$ subgroup governing the mixing between second- and third-generation left-handed leptons. We derive constraints on the corresponding spurion parameter $\delta$ from current data on lepton flavor violating (LFV) and lepton flavor universality (LFU) observables, finding that $R_{K^{(*)}}$ and $\mathcal{B}(B_s \to \mu \mu)$ give the most stringent bound on $\delta$, yielding ${|\delta|<0.051}$ at~95\%~CL. In addition, we provide updated bounds for LFV decay rates and discuss prospects for future sensitivity improvements, finding that future LFV searches could further tighten constraints on the mixing between second- and third-generation leptons.
\end{abstract}




\end{frontmatter}




\section{Introduction}
\label{introduction}

The explanation of the Higgs hierarchy problem calls for new physics (NP) at the TeV scale to stabilize the Higgs mass. However, stringent constraints from both direct and indirect searches for NP suggest that the new scale must be higher, unless NP exhibits a rather specific flavor structure. In particular, the absence of significant deviations from Standard Model (SM) predictions in flavor-changing processes implies that NP couplings to the light fermion families must be nearly degenerate.
Moreover, bounds from direct searches impose strong constraints on new states coupled to light families.

As shown in~\cite{Allwicher:2023shc}, the hypothesis of NP at the TeV scale that couples predominantly to third-generation fermions, with weaker and quasi-degenerate interactions with the lighter generations, characterized by an approximate $U(2)^5$ flavor symmetry, is fully consistent with constraints from direct searches, electroweak precision observables, and flavor observables. 
These assumptions on new physics are present in a wide class of compelling new physics models. Indeed, the $U(2)^5$ flavor symmetry has been discussed in the context of supersymmetric models~\cite{Barbieri:2011ci,Larsen:2012rq, Papucci:2011wy}, composite models~\cite{Barbieri:2012tu, Matsedonskyi:2014iha, Panico:2016ull, Covone:2024elw}, and more recently, in flavor non-universal gauge models~\cite{Bordone:2017bld, FernandezNavarro:2023rhv, Davighi:2023evx, Davighi:2023iks, Greljo:2018tuh, Fuentes-Martin:2022xnb}.

While the $U(2)^5$ flavor symmetry is very efficient in addressing direct and indirect NP constraints, and is a very good starting point to describe the Yukawa sector of the SM, it cannot be an exact symmetry. 
The $U(2)^5$ symmetry indeed forbids light fermion masses and the mixing between the third and the light generations. In the quark sector, 
hints of $U(2)^5$ breaking are also provided by a series of deviations from the SM predictions observed in rare and semileptonic $B$ decays.
The best description of current data is obtained under the hypothesis that the leading $U(2)^5$ breaking, both in the SM sector and beyond, is the breaking of $U(2)_q$, which is responsible for the third-light mixing in the Cabibbo--Kobayashi--Maskawa (CKM) mixing matrix~\cite{Allwicher:2024ncl}. 
The situation of the lepton sector is less clear. The goal of this paper is to clarify this aspect, investigating in particular if and how large the breaking can be in the $U(2)_\ell$ direction, related to the mixing of left-handed leptons.

Under the assumption of decoupling heavy new physics, the natural framework to systematically address these questions is the so-called Standard Model Effective Field Theory (SMEFT) (see the recent review~\cite{Isidori:2023pyp}). Within the SMEFT, the effects of new physics are encoded in the Wilson coefficients (WCs) of a tower of higher-dimensional operators built in terms of SM fields. A $U(2)^5$ flavor symmetry acting on the lightest two SM families is then imposed on the SMEFT operators~\cite{Barbieri:2012uh, Isidori:2012ts}. These assumptions significantly reduce the number of relevant operators and, consequently, free parameters.
The formalism also allows to systematically describe breaking terms introducing (small) spurion terms with well-defined transformation properties under the flavor symmetry.

In this work, we build upon the SMEFT framework of~\cite{Allwicher:2024ncl}, where current data is analyzed allowing non-vanishing WCs for semileptonic operators 
built in terms of third-generation fields and 
a single $U(2)_q$--breaking spurion.
We aim to extend that analysis by introducing an additional spurion that breaks the $U(2)_\ell$ symmetry.
This allows us to assess the degree to which such a breaking is compatible with current experimental data, and to explore the resulting implications for future measurements, with particular emphasis on Lepton Flavor Violating (LFV) processes and Lepton Flavor Universality (LFU) tests.

The rest of the paper is organized as follows. In Section~\ref{sec:eftframework}, we describe our effective field theory (EFT) setup, including the renormalization group evolution (RGE) and matching procedures. In Section~\ref{sec:numanalysis} we describe our determination of the relevant parameter representing the $U(2)_\ell$ breaking. We also discuss the future prospects of different experimental channels that can probe the breaking of $U(2)_\ell$, including a series of predictions on their upper limits, as implied by our fit to the $U(2)_\ell$--breaking parameter. Finally, in Section~\ref{sec:conclusions} we discuss the main conclusions that can be drawn from our work. Additionally, we list some relevant theoretical expressions in~\ref{app:ObsTh} and~\ref{app:Wilsons}.

\section{EFT framework}
\label{sec:eftframework}
As already motivated in the Introduction, the EFT framework adopted in this work is constructed under the assumption of heavy NP predominantly coupled to third-generation fermions, following an approximate symmetry 
\begin{equation}
    U(2)^5 = U(2)_q \times U(2)_u \times U(2)_d \times U(2)_\ell \times U(2)_e \;.
\end{equation}
This flavor symmetry distinguishes the light generations of each fermion field, which transform as a doublet, and the heavy one, which is a singlet. These UV assumptions are imposed at the level of the SMEFT, where all heavy degrees of freedom beyond the Standard Model (BSM) are integrated out, and impact the theory only via higher-dimension operators. As the leading effect in the $1/\Lambda_{\rm NP}$ expansion, we thus consider dimension-six SMEFT operators built in terms of a minimal set of spurions that break the $U(2)^5$ symmetry along the $U(2)_q$ and $U(2)_\ell$ directions. These spurions, responsible for the heavy-to-light mixing in the quark and lepton Yukawa couplings, are parametrized as 
\begin{equation}
    \tilde{V}_q^i =  \left(
    \begin{array}{c}
        - \varepsilon V_{td} \\
        - \varepsilon V_{ts} 
    \end{array} \right) \;, \qquad
    \tilde{V}_\ell^\alpha = \left(
    \begin{array}{c}
        \delta \sin{\theta_e} \\
        \delta \cos{\theta_e} 
    \end{array} \right) \;,
    \label{eq:Vlgen}
\end{equation}
where $\epsilon$ and $\delta$ are assumed to be real parameters. 
It follows from these definitions that $|\tilde V_\ell| = \delta$.

Being aligned to the heavy-light mixing 
terms in the CKM matrix, $\tilde{V}_q$ 
breaks $U(2)_q$ in a minimal way
(i.e.~it is the same spurion present in the quark Yukawa couplings). Beside minimality,
this choice is motivated by the lack of 
 evidence for non-minimal breaking in the quark sector~\cite{Allwicher:2024ncl}\footnote{The parameter $\kappa$, characterizing the departure from minimal breaking, remains loosely constrained and compatible with ${\kappa = 1}$ in their analysis.}. 
The form of $\tilde{V}_\ell$ 
in Eq.~(\ref{eq:Vlgen}) corresponds to a generic direction in $U(2)_\ell$ space,
characterized by the angle $\theta_e$. Given the tightness of the bounds on LFV processes involving electrons (see e.g.~\cite{Asadi:2025dii}), in the following we set $\cos{\theta_e} = 1 ~(\sin{\theta_e}=0)$, and therefore restrict our analysis to LFV processes involving third- and second-generation leptons\footnote{Allowing for mixing angles as small as $\theta_e \approx 0.1$ would already push our bound on $\delta$ to values more than $10$ times smaller, and at an equal electron-muon mixing, $\theta_e \approx \pi/4$, it would be pushed down to values more than $150$ times smaller.}. 

Regarding the ambiguity in what constitutes the third generation of left-handed quarks, we choose a down-aligned basis, where the quark doublets are
\begin{equation}
    q_L^i = \left(
    \begin{array}{c}
        V_{ui} u_L + V_{ci} c_L + V_{ti} t_L \\
        d^i_L
    \end{array} \right) \;,
\end{equation}
with $u_L, d_L$ being the quark mass eigenstates, such that $q_L^3$ is a $U(2)_q$ singlet. As for the leptons, there is no ambiguity within the SM: $\ell_L^3 = (\nu_\tau,\tau_L)$. 

With these assumptions, the relevant part of the SMEFT Lagrangian is
\begin{equation}
\label{eq:LagrangianSMEFT}
\begin{aligned}
    \mathcal{L}_{\rm SMEFT}^{\rm NP} \; \supset \; & \frac{1}{\Lambda_{\rm NP}^2}\left[ C^+_{\ell q} \right]_{\alpha\beta ij} \left[Q^+_{\ell q}\right]_{\alpha\beta ij} 
   \\ & + \frac{1}{\Lambda_{\rm NP}^2} \left[ C^-_{\ell q} \right]_{\alpha\beta ij} \left[Q^-_{\ell q}\right]_{\alpha\beta ij}\;,
\end{aligned}
\end{equation}
where, from now on, we will assume $\Lambda_{\rm NP} = 1$ TeV
, and
\begin{align}
    \left[Q^{\pm}_{\ell q}\right]^{\alpha \beta i j} =\;& (\bar{q}_L^i \gamma^\mu q_L^j)(\bar{\ell}_L^\alpha \gamma_\mu \ell_L^\beta) 
    \\ & \pm (\bar{q}_L^i \gamma^\mu \sigma^a q_L^j)(\bar{\ell}_L^\alpha \gamma_\mu \sigma^a \ell_L^\beta) \;.
\end{align}
This defines to a UV setup equivalent to that of~\cite{Allwicher:2024ncl}, including the choice of $\Lambda_{\rm NP}$ and of a down-aligned basis\footnote{Still, this mostly corresponds to the conservative approach, as seen e.g. in~\cite{Allwicher:2023shc}, where an up-aligned basis generally leads to stronger constraints on the SMEFT coefficients considered in our analysis.}, barring the introduction of an additional spurion for the lepton sector. Also for consistency with \cite{Allwicher:2024ncl}, we adopt the simplifying hypothesis of a rank-one alignment in the NP structure~\cite{Marzocca:2024hua}. Under this assumption, all flavor misalignment from the third generation arises exclusively from the spurions.

This flavor setup yields the following scaling of the relevant SMEFT coefficients:
\begin{gather}
\label{eq:SMEFTscaling}
    \left[ C^\pm_{\ell q} \right]_{\alpha\beta ij} = C^\pm_{\ell q} \, (\mathcal{V}_\ell^\dagger)_\alpha \, (\mathcal{V}^{\vphantom{\dagger}}_\ell)_{\beta} \, (\mathcal{V}_q^\dagger)_i \, (\mathcal{V}^{\vphantom{\dagger}}_q)_{j} \;, \\
    \text{with} \quad \mathcal{V}_{\ell,q} = \left(
    \begin{array}{c}
         \tilde V_{\ell,q} \\ 
         1
    \end{array} \right) \;,
\end{gather}
and where the parameters $C^{\pm}_{\ell q}$ govern the overall size of the mainly third-generation NP. Let us also point out that we neglect scalar operators in the Lagrangian of Eq.~(\ref{eq:LagrangianSMEFT}). This is justified by the results of the analysis in~\cite{Allwicher:2024ncl}, which showed that the corresponding WCs are mostly irrelevant, and setting them to zero is well supported by data.

The EFT framework resulting from our flavor assumptions is then governed by only four independent parameters: $C^+_{\ell q}, C^-_{\ell q}, \epsilon$ and $\delta$.

We would like to point out that our setup does not intend to explain the structure of the neutrino mass matrix, which appears to be anarchic compared to the quark and charged-lepton mass matrices (see e.g. the review in~\cite{King:2014nza}). In this sense, the hierarchical structures introduced in the $U(2)^5$ framework can pose a problem to model-building efforts based on flavor deconstruction. However, as shown in \cite{Greljo:2024ovt}, neutrino anarchy may still arise in minimal models involving right-handed neutrinos by imposing that the latter are not singlets of the extended gauge group. This allows for a cancellation of all hierarchy among the active neutrinos, while producing the correct charged-fermion mass and mixing hierarchies. 
Because of this, and given that our setup will produce no sizable mixing with the dimension-five Weinberg operator $Q_{\ell\ell\varphi\varphi}$, we will refrain from including any neutrino flavor-violating processes in our analysis.

\subsection{RGE and Matching}
To study the low-energy phenomenological implications of our UV setup, we use the framework of the Low-Energy EFT below the EW scale (LEFT)~\cite{Jenkins:2017jig,Jenkins:2017dyc}. The NP effects on the Wilson coefficients of the LEFT can be obtained by calculating the renormalization group (RG) evolution of the SMEFT coefficients from the NP scale (${\Lambda_{\rm NP} = 1 \text{ TeV}}$) down to the EW scale (${\Lambda_{\rm EW} \sim M_Z \approx 91.2 \text{ GeV}}$), then calculating their matching to the LEFT, and then evolving the LEFT coefficients down to the low-energy scale of the observables (${\Lambda_{B} \sim m_b \approx 4.8 \text{ GeV}}$). Given that the Weinberg operator, irrelevant for our study, is the only dimension-five operator in the SMEFT, and that our high-energy setup leads to negligible double-dipole insertions within the LEFT, we limit ourselves to considering single insertions of effective operators in the running. The RGE can then be solved exactly in terms of an evolution matrix in either EFT~\cite{Fuentes-Martin:2020zaz},
\begin{equation}
    \mathcal{C}_i(\mu) = U^{\mathcal{C}}_{ij}(\mu,\mu_0) \, \mathcal{C}_j(\mu_0) \;.
\end{equation}
Concerning the matching between the SMEFT ($C_i$) and the LEFT ($L_i$), it can also be written in terms of a single matrix upon neglecting double insertions,
\begin{equation}
    L_i(\mu_m) = L_i^{\rm (SM)}(\mu_m) + M_{ij} \, C_j(\mu_m) \;,
\end{equation}
where the SM matching $L_i^{\rm (SM)}(\mu_m)$ is computed in absence of NP ($C_i=0 \; \forall \;i$). Let us note here that this implicitly neglects the small effect of higher-dimension Wilson coefficients in the running of the SM parameters which, in principle, could have an impact on the SM matching. We can therefore write the whole evolution process described above, from $\Lambda_{\rm NP}$ to $\mu_{\rm low}$, in terms of a couple of evolution matrices:
\begin{equation}
    L_i(\mu_{\rm low}) =\; L_i^{\rm (SM)}(\mu_{\rm low}) + \mathcal{U}_{ij}(\mu_{\rm low},\mu_m,\Lambda_{\rm NP}) \, C_j(\Lambda_{\rm NP}) \;.
\end{equation}
where we have defined
\begin{gather}
    L_i^{\rm (SM)} \,=\, U^L_{ik}(\mu_{\rm low},\mu_m) \, L_k^{\rm (SM)}(\mu_m) \;,
    \\[1mm]
    \mathcal{U}_{ij} \,=\, U^L_{ik}(\mu_{\rm low},\mu_m) \, M_{kl}(\mu_m) \, U^C_{lj}(\Lambda_{\rm EW},\Lambda_{\rm NP}).
\end{gather}
The consistent approach to the perturbative expansion of the evolution matrix $\,\mathcal{U}_{ij}$ corresponds, at leading order (LO), to combining the LL RGE and tree-level matching. These are fully available for the JMS basis of the LEFT and the Warsaw basis of the SMEFT, up to dimension six, in \texttt{DsixTools}~\cite{Fuentes-Martin:2020zaz}, for instance. The NLO evolution, corresponding to the combination of the NLL RGE and the one-loop matching, is not yet fully available in the literature. While the one-loop matching is fully known from~\cite{Dekens:2019ept}, and the two-loop RGE in the LEFT is known in most of its two-loop contributions~\cite{Naterop:2024cfx,Aebischer:2025hsx,Naterop:2025lzc,Naterop:2025cwg}, the two-loop RGE in the SMEFT is not yet available in the literature. 

As such, we choose to take a hybrid approach for $\,\mathcal{U}_{ij}$, by which we combine the LL RGE and one-loop matching using \texttt{DsixTools}. We list the low-energy (${\mu = 5 \text{ GeV}}$) expressions of the Wilson coefficients that are relevant to our analysis resulting from our implementation of the RGE and matching in~\ref{app:Wilsons}.

Given that, for most observables, we limit ourselves to the calculation of tree-level amplitudes, which cannot cancel the dependence on the renormalization scheme of the finite terms in the one-loop matching, our results will ultimately suffer from some degree of scheme-dependence~\cite{Herrlich:1994kh}. Still, we find this more desirable than neglecting the one-loop matching altogether, and rather expect the scheme-dependent terms to have no sizable impact in our analysis\footnote{We expect the Wilson coefficients whose matching is dominated by the one-loop terms to yield the biggest sensitivity to the renormalization scheme. This includes the coefficients of all lepton dipoles and four-lepton operators, as well as the light-quark $\bar\mu \tau \bar q q$ operators ($q=u,d,s$). Still, all of these contribute to the likelihood via upper limits, and thus we expect this scheme-dependence to have little impact.}.
Regarding uncertainties in the evolution matrix $\,\mathcal{U}_{ij}$, we estimate them by varying both the low-energy scale and the matching scale, the former in the standard factors of $1/2$ and $2$ around a central value of $5$ GeV, and the latter in $\mu = [70,120]$ GeV (the central value being $91.2$ GeV). What we observe is that these variations can provide relative differences of up to $10\%$ (accounted for almost exclusively by the variations in the matching scale) in the NP contributions to most of the relevant Wilson coefficients. In order not to overly complicate the setup, we incorporate this uncertainty by repeating the statistical analysis (see the following section) for $\mu_m = 70,120$ GeV, and for $\mu_{\rm low} = 2, 5, 10 \text{ GeV}$, and studying the resulting dependence.

\section{Numerical Analysis}
\label{sec:numanalysis}

The set of observables we find more promising to study within the framework described above is listed in Table~\ref{tab:observables}. 
It includes, for instance, the branching ratios of several LFV meson and tau decays. The experimental bounds for these decay channels are listed mostly as in~\cite{Plakias:2023esq}\footnote{For the experimental limits that we take from the PDG~\cite{ParticleDataGroup:2024cfk}, we transform from 90\% to 95\% CL assuming a Gaussian distribution.}, where they are extracted from~\cite{LHCb:2017hag, LHCb:2019bix, LHCb:2022lrd, LHCb:2022wrs, Belle:2022pcr, Belle:2023jwr, ParticleDataGroup:2020ssz} and transformed into $95\%$ CL bounds according to~\cite{Calibbi:2017uvl}. Among them, we include the branching ratios for the LFV quarkonium decays ${J/\psi\to\mu\tau}$ and ${\Upsilon\to\mu\tau}$ since they have been pointed out as interesting alternative probes for LFV~\cite{Hazard:2016fnc,Calibbi_2022},  although their bounds are not as low as for $B$-meson or tau decays. These channels are particularly noteworthy as they realize lepton-flavor violation without involving quark-flavor transitions, thus offering a distinct flavor structure. Moreover, they can potentially induce non-negligible third-to-light quark family mixing through loop-level effects.

\begin{table}[t]
    \centering
    \renewcommand{\arraystretch}{1.3}
    \begin{tabular}{c|c|c}
    Observable & Experimental Bound/Measurement & Ref. \\
     \hline
       $\mathcal{B}(\tau \to \mu \gamma)$  &  $<5.6 \times 10^{-8}$ & \cite{Plakias:2023esq} \\ 
       $\mathcal{B}(\tau \to \mu e e)$  &  $<2.4 \times 10^{-8}$ & \cite{Plakias:2023esq} \\
       $\mathcal{B}(\tau \to \mu \mu \mu)$  &  $<2.8 \times 10^{-8}$ & \cite{Plakias:2023esq} \\ 
       $\mathcal{B}(\tau \to \mu \rho)$ & $<1.6 \times 10^{-8}$ & \cite{Plakias:2023esq} \\ 
       $\mathcal{B}(\tau \to \mu \phi)$  &  $<2.9 \times 10^{-8}$ & \cite{ParticleDataGroup:2024cfk} \\ 
       $\mathcal{B}(J/\psi \to \mu \tau)$  &  $<2.7 \times 10^{-6}$ & \cite{Plakias:2023esq} \\ 
       $\mathcal{B}(\Upsilon \to \mu \tau)$  &  $<3.6 \times 10^{-6}$ & \cite{Plakias:2023esq} \\ 
       $\mathcal{B}(B_s \to \mu\tau)$  &  $<4.2 \times 10^{-5}$ & \cite{Plakias:2023esq} \\ 
       $\mathcal{B}(B \to K \mu\tau)$  &  $<4.1 \times 10^{-5}$ & \cite{Plakias:2023esq} \\ 
       $\mathcal{B}(B \to K^* \mu\tau)$  &  $<2.2 \times 10^{-5}$ & \cite{Plakias:2023esq} \\  
       $\mathcal{B}(B_s \to \phi \mu\tau)$  &  $<1.3 \times 10^{-5}$ & \cite{ParticleDataGroup:2024cfk} \\ \hline
       $\mathcal{B}(B_s \to \tau\tau)$  &  $<6.8 \times 10^{-3}$ & \cite{ParticleDataGroup:2024cfk} \\ 
       $\mathcal{B}(B_s \to \mu\mu)$  &  $(3.34 \pm 0.27) \times 10^{-9}$ & \cite{ParticleDataGroup:2024cfk}\\ \hline
       $R_K[0.1, 1.1]$ & $0.994^{+0.090}_{-0.082} ~(\rm{stat})^{+0.029}_{-0.027} ~(\rm{syst}) $ & \cite{LHCb:2022qnv}\\
       $R_K[1.1,6]$ & $0.949^{+0.042}_{-0.041} ~(\rm{stat})^{+0.023}_{-0.023} ~(\rm{syst}) $ & \cite{LHCb:2022qnv}\\
       $R_K[14.3,22.9]$ & $1.08^{+0.11}_{+0.04} ~(\rm{stat})^{+0.09}_{-0.04} ~(\rm{syst}) $ & \cite{LHCb:2025ilq}\\
       $R_{K^*}[0.1, 1.1]$ & $0.927^{+0.093}_{-0.087} ~(\rm{stat})^{+0.034}_{-0.033} ~(\rm{syst}) $ & \cite{LHCb:2022qnv}\\
       $R_{K^*}[1.1, 6]$ & $1.027^{+0.072}_{-0.068} ~(\rm{stat})^{+0.027}_{-0.027} ~(\rm{syst}) $ & \cite{LHCb:2022qnv}
    \end{tabular}
    \caption{Experimental bounds/measurements on the relevant observables. All the bounds are given at 95\% CL. Whenever a decay has two different leptons in the final state, we quote the sum over the two particle-antiparticle possibilities, i.e. $\mu \tau \equiv \mu^+ \tau^- \oplus \mu^- \tau^+$.}
    \label{tab:expbounds}
    \label{tab:observables}
\end{table}

We decide to include the branching ratio of ${\tau \to \mu e e}$, even though it involves first-generation leptons, because the corresponding Wilson coefficients in the LEFT are generated at loop-level in the matching~\cite{Dekens:2019ept}, just as the ones mediating ${\tau \to \mu \mu \mu}$. Furthermore, this picture is mostly insensitive to the choice of the angle $\theta_e$. 

We also include the branching ratio of ${B_s\to \mu\mu}$~\cite{ParticleDataGroup:2024cfk} and the LFU ratios $R_{K^{(*)}}$~\cite{LHCb:2022qnv} which, albeit not featuring explicit LFV, are sensitive to the spurion parameter $\delta$ due to the flavor structure of our setup. See \ref{lfuapp}, \ref{bsllapp} for further details. Regarding the branching ratios of $B \to K^{(*)}\mu\mu$ and the angular observables of $B \to K^*\mu \mu$, the shift in the Wilson coefficient $C_9$ (see e.g. \cite{Isidori:2025dkp, Bordone:2024hui}) necessary to explain the experimental tensions with the SM is already accounted for by the analysis in \cite{Allwicher:2024ncl}. Furthermore, since these observables are less clean than the LFU ratios, we do not expect them to give stronger constraints on $\delta$ and therefore we do not re-analyze them.

Finally, we also quote the experimental bound on ${B_s\to \tau\tau}$, which we shall use to study the impact of our EFT framework on $b \to s(d) \tau \tau$ transitions. Given that these processes do not influence our determination of $\delta$, we refrain from adding any more of them in Table~\ref{tab:expbounds}.

For most observables, we limit ourselves to the calculation of the tree-level contributions to the amplitude, leaving the RGE and matching calculations as the only loop-level effects we consider. This being said, we do include the one-loop contribution from the four-lepton operators to the branching ratio of $\tau \to \mu \gamma$, given that it is comparable in size to the tree-level contributions from the dipole operators.

Table~\ref{tab:scaling} summarizes the leading dependence on the parameters in our framework, both in the SMEFT and in the LEFT, of the amplitudes for a few different classes of transitions. 
It goes without saying that, after matching and RGE evolution down to the low-energy scale, the picture gets shifted from the naive expectation at the high-energy scale. In the Table, we do not include non-hadronic transitions, such as $\tau \to \mu \gamma$ or $\tau \to \mu \ell \ell$, given that their leading contributions only appear at one-loop order in our SMEFT setup. For the complete expressions of both the observables and Wilson coefficients, we refer the reader to~\ref{app:ObsTh} and~\ref{app:Wilsons}, respectively.

\begin{table}[t]
    \centering
    \renewcommand{\arraystretch}{1.3}
    \begin{tabular}{c|c|c|c}
    Transition & SMEFT Scaling & \; LEFT Scaling \; \\
     \hline
       $b \to s \tau \tau$ & $C_{\ell q}^+ \, V_{ts} \, \epsilon$ & $C_{\ell q}^+ \, V_{ts} \, \epsilon$ \\
       $b \to s \mu \mu$ & $C_{\ell q}^+ \, V_{ts} \, \epsilon \,c_e^2 \, \delta^2$ & $C_{\ell q}^+ \, V_{ts} \, \epsilon \, (0.005 - c_e^2 \, \delta^2)$ \\
       $b \to s e e$ & $C_{\ell q}^+ \, V_{ts} \, \epsilon \,s_e^2 \, \delta^2$ & $C_{\ell q}^+ \, V_{ts} \, \epsilon \, (0.005 - s_e^2 \, \delta^2)$ \\
       $b \to s \tau \mu$ & $C_{\ell q}^+ \, V_{ts} \, \epsilon \,c_e \, \delta$ & $C_{\ell q}^+ \, V_{ts} \, \epsilon \,c_e \, \delta$ \\
       $b \bar{b} \to \tau \mu$ & $C_{\ell q}^+ \,c_e \, \delta$ & $(C_{\ell q}^+ + 0.05 \,C_{\ell q}^-) \,c_e \, \delta$ \\
       $c \bar{c} \to \tau \mu$ & $C_{\ell q}^- \,|V_{ts}|^2\epsilon^2 \,c_e \, \delta$ & $(C_{\ell q}^+ - 0.7 \,C_{\ell q}^-) \,c_e \, \delta$ \\
       $\tau \to \mu s s$ & $C_{\ell q}^+ \,|V_{ts}|^2\epsilon^2 \,c_e \, \delta$ & $(C_{\ell q}^+ - 0.7 \,C_{\ell q}^-) \,c_e \, \delta$ \\
       $\tau \to \mu dd$ & $C_{\ell q}^+ \,|V_{td}|^2\epsilon^2 \,c_e \, \delta$ & $(C_{\ell q}^+ - 0.7 \,C_{\ell q}^-) \,c_e \, \delta$ \\
       $\tau \to \mu uu$ & $C_{\ell q}^- \,|V_{td}|^2\epsilon^2 \,c_e \, \delta$ & $(C_{\ell q}^+ - 0.7 \,C_{\ell q}^-) \,c_e \, \delta$ 
       
    \end{tabular}
    \caption{Scaling of the leading NP contributions of a few types of transition amplitudes in the SMEFT at the high-energy scale, according to Eq.~(\ref{eq:SMEFTscaling}), and in LEFT after evolving down to low energies, which presents clear deviations from the naive SMEFT scaling in many cases. We refer to~\ref{app:Wilsons} for the explicit expressions of the low-energy Wilson coefficients.}
    \label{tab:scaling}
\end{table}

We list the set of inputs most relevant to our analysis in Table~\ref{tab:SMvalues}. 
The SM central values for the LFU ratios $R_{K^{(*)}}$ are quoted from \texttt{Flavio}~\cite{Straub:2018kue}, while their uncertainties are expected to be at the percent level according, for instance, to \cite{Bordone:2016gaq,Isidori:2020acz}.
In the case of $\mathcal{B}(B_s \to \mu^+\mu^-)$, the SM prediction is taken from \cite{Czaja:2024the}. 

Concerning the CKM matrix, we follow the same prescription as in \cite{Allwicher:2024ncl}. Given the tension in the determination of $|V_{cb}|$, this parameter is extracted from a combination of the global fit for the inclusive determination~\cite{Finauri:2023kte} and the recent exclusive determination~\cite{Bordone:2024weh}, which also agrees with the previous analyses in \cite{Martinelli:2021onb,Martinelli:2021myh,Martinelli:2022xir,Martinelli:2023fwm} and with the global fits in \cite{UTfit:2022hsi,FlavourLatticeAveragingGroupFLAG:2024oxs}. The Wolfenstein parameters $\bar\rho$ and $\bar\eta$ are fixed from observables sensitive to the angles only, and $\lambda$ from super-allowed $\beta$ decays. The value of $|V_{cb}|$ obtained as described above is then used to extract the parameter $A$. This prescription yields the values in Table~\ref{tab:SMvalues}.

\begin{table}[t]
\renewcommand{\arraystretch}{1.3}
    \centering
    \begin{tabular}{c|c|c}
    Input Parameter & Value & Ref.\\ \hline
        $\mathcal{B}(B_s \to \mu \mu)_{\rm SM}$ & $(3.64 \pm 0.12) \times 10^{-9}$ & \cite{Czaja:2024the} \\
        $R_K [0.1, 1.1]_{\rm SM}$ & $0.99\pm 0.01$ & \cite{Straub:2018kue,Bordone:2016gaq,Isidori:2020acz} \\
        $R_K [1.1,6]_{\rm SM}$ & $1.00 \pm 0.01$ & \cite{Straub:2018kue,Bordone:2016gaq,Isidori:2020acz} \\
        $R_K [14.3,22.9]_{\rm SM}$ & $1.00 \pm 0.01$ & \cite{Straub:2018kue,Bordone:2016gaq,Isidori:2020acz} \\
        $R_{K^*}[0.1, 1.1]_{\rm SM}$ & $0.983 \pm 0.014$ & \cite{Bordone:2016gaq} \\
        $R_{K^*}[1.1, 6]_{\rm SM}$ & $1.00 \pm 0.01$ & \cite{Straub:2018kue,Bordone:2016gaq} \\
        \hline
        $A$ &  $0.816 \pm 0.017$ & \cite{Allwicher:2024ncl} \\
        $\lambda$ & $0.2251 \pm 0.0008$ & \cite{Allwicher:2024ncl} \\
        $\bar\rho $ & $0.144 \pm 0.016$ & \cite{Allwicher:2024ncl} \\
        $\bar\eta$ & $0.343 \pm 0.012$ & \cite{Allwicher:2024ncl} 
    \end{tabular}
    \caption{Input parameters relevant to our analysis. 
    }
    \label{tab:SMvalues}
\end{table}

\subsection{Results}
All observables prove to be somewhat competitive in constraining the LFV parameter $\delta$, providing bounds of $O(1)$ or below when studied separately, except for the quarkonium LFV decays. This is especially the case of $J/\psi\to\mu\tau$, which provides a much looser constraint. We clearly identify the central-$q^2$ bins of $R_{K^{(*)}}$ and the high-$q^2$ bin ($q^2 >14.3$ GeV) of $R_K$, together with the branching ratio of $B_s\to\mu^+\mu^-$ as the ones imposing the most stringent bounds on $\delta$. We therefore choose this set of four observables for our fit.
Given that we want to build upon the work done in~\cite{Allwicher:2024ncl}, we use the chi-square function resulting from their fit to the parameters $C_{\ell q}^+, C_{\ell q}^-$ and $\epsilon$ as a nuisance contribution ($\chi^2_{\rm nuis}$) to add to the chi-square function of our set of observables,  $\chi^2_{\text{obs}}$ . 
For that purpose, we report here their central values (best-fit points) and covariance matrix\footnote{We thank the authors of \cite{Allwicher:2024ncl} for providing the full chi-square function resulting from their analysis.
},
\begin{gather}
    C_{\ell q}^+ \big|_{\rm bf} \approx -0.406 \;, \quad C_{\ell q}^- \big|_{\rm bf} \approx 0.0943 \;, \quad \varepsilon \big|_{\rm bf} \approx 2.90 \;,
    \\[2mm]
    \Sigma_{ij} = \left(
    \begin{array}{ccc}
        13.7 & -0.382 & 11.0 \\
        -0.382 & 1.08 & -12.1 \\
        11.0 & -12.1 & 213 
    \end{array} \right) \times 10^{-3} \;.
\end{gather}
The relative uncertainty in the determination of the nuisance parameters ${\{C_{\ell q}^+, C_{\ell q}^-,\epsilon\}}$ turns out to be the dominant source of uncertainty, much larger than that of most other input parameters. We therefore neglect the latter and, for the statistical analysis, consider the nuisances as the only source of uncertainty in the input parameters, aside from that of the SM predictions in Table~\ref{tab:SMvalues}.

In order to fit the spurion parameter $\delta$, we assume a Gaussian distribution to determine the full likelihood, 
\begin{equation}
    L(\vec{\theta}) = \mathcal{N} \, \exp\Big[-\frac{1}{2}(\chi^2_{\rm obs} + \chi^2_{\rm nuis})\Big] \;,
\end{equation}
where $\vec{\theta} = \{C_{lq}^+, C_{lq}^-, \epsilon, \delta\}$,
and then sample over it using Monte-Carlo methods\footnote{We choose this over profiling because the profiled likelihood becomes somewhat flat for values of ${|\delta| > 0.1}$, which is problematic for the determination of the proper CL regions.}. 
Our result on the allowed region for $\delta$ reads
\begin{equation}
\label{eq:mainbound}
    |\delta| < 0.051 \quad ({\rm at}\; 95\% \; {\rm CL}) \;,
\end{equation}
with ${\chi^2/{\rm d.o.f.} \approx 0.94}$ for the best-fit point. 
Incidentally enough, this value is comparable to ${|V_{ts}| \sim 0.04}$, which is the natural size of the $U(2)_q$--breaking spurion inferred from quark Yukawa couplings.

Following the discussion at the end of Section~\ref{sec:eftframework}, we have repeated the analysis for the different values of the matching scale $\mu_{\rm m}$ and $\mu_{\rm low}$, and found numerically equivalent bounds in all cases, thus making the bound in Eq.~(\ref{eq:mainbound}) largely independent of the choice of renormalization scale.     

\begin{figure}[t]
    \centering
\includegraphics[width=0.95\linewidth]{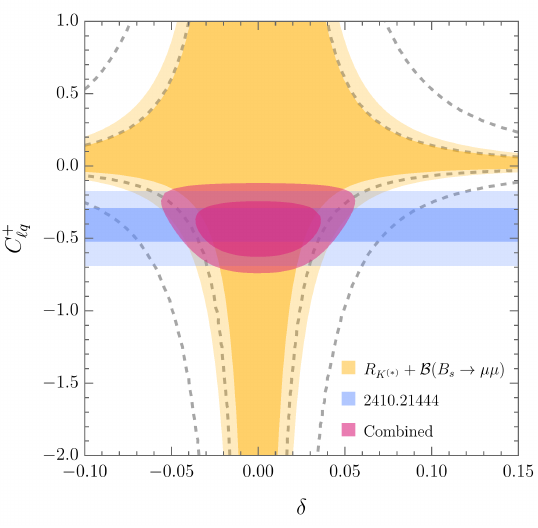}
    \caption{Main results of our fit in the $\{\delta,C_{\ell q}^+\}$ plane. The $1\sigma$ and $2\sigma$ regions from the results in~\cite{Allwicher:2024ncl} provide horizontal bands (blue). The $1\sigma$ and $2\sigma$ bounds imposed by $\mathcal{B}(B_s\to\mu\mu)$ and $R_{K^{(*)}}$, with $C_{\ell q}^-$ and $\epsilon$ fixed to their best-fit values, provide a cross-shaped region (orange). The gray dashed contours represent the $2\sigma$ CL regions with fixed $\epsilon = 1.0$ (inner), and $\epsilon=3.6$ (outer). The $1\sigma$ and $2\sigma$ regions from the combined fit yield the central region in pink.}
    \label{fig:Cp-Delta}
\end{figure}

In Figure~\ref{fig:Cp-Delta} we show the overall results of the fit in a $\{\delta, C_{\ell q}^+ \}$ two-dimensional plane. The results of~\cite{Allwicher:2024ncl} provide a bound on $C_{\ell q}^+$ represented by the horizontal blue band. The projection of the bounds provided by our chosen set of observables, $R_{K^{(*)}}$ and $\mathcal{B}(B_s\to\mu\mu)$, over the slice corresponding to the best-fit values of $C_{\ell q}^-$ and $\epsilon$, provides the orange cross-shaped region. The combined $1 \sigma$ and $2\sigma$ regions are shown in pink. While the overall dependence on $C_{\ell q}^-$ of this region is largely negligible, we represent the dependence on $\epsilon$ by including the contours of the $2\sigma$ region (dashed lines) for two additional representative values at the limits of the marginalized $2\sigma$ CL region of $\epsilon$ ($\epsilon = 1.0$ and $\epsilon = 3.6$). In this respect, we see that, even if $\epsilon$ is not too tightly constrained, the picture of the constraints on $\delta$ remains mostly the same.                        

\begin{figure}[t]
    \centering
\includegraphics[width=0.95\linewidth]{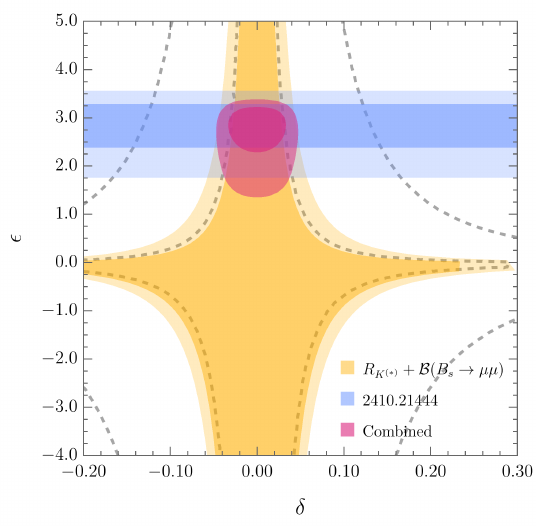}
    \caption{Main results of our fit in the $\{\delta,\epsilon\}$ plane. The $1\sigma$ and $2\sigma$ region from the results in~\cite{Allwicher:2024ncl} provide horizontal bands (blue). The $1\sigma$ and $2\sigma$ bounds imposed by $\mathcal{B}(B_s\to\mu\mu)$ and $R_{K^{(*)}}$, with $C_{\ell q}^\pm$ fixed to their best-fit values, provide a cross-shaped region (orange). The gray dashed contours represent the $2\sigma$ CL regions with fixed $C_{\ell q} = -0.77$ (inner), and $C_{\ell q} = -0.05$ (outer). The $1\sigma$ and $2\sigma$ regions from the combined fit yield the central region in pink.}
    \label{fig:Eps-Delta}
\end{figure}
Figure~\ref{fig:Eps-Delta} depicts an analogous plot in the $\{\delta, \epsilon \}$ two-dimensional plane. The additional dashed lines represent the $2\sigma$ CL regions for two additional representative values at the limits of the marginalized $2\sigma$ CL region of $C_{\ell q}^+$ ($C_{\ell q}^+ = -0.77$ and $C_{\ell q}^+ = -0.05$). Given the stronger dependence of the likelihood on $C_{\ell q}^+$, these additional bands cover a much wider region of $\delta$ in this case.

Another interesting quantity we can probe in our framework is the ratio between the spurions $|\tilde V_\ell|/|\tilde V_q|$, which characterizes the relative difference between the magnitude of the breaking of $U(2)_\ell$ and $U(2)_q$. For this ratio we can also provide an upper limit,
\begin{equation}
    |\tilde V_\ell|/|\tilde V_q| < 0.69 \quad ({\rm at}\; 95\% \; {\rm CL}) \;.
\end{equation}
Even though this shows that current data implies $|\tilde V_\ell| \lesssim |\tilde V_q|$, it still falls short in terms of resolving whether they are of the same magnitude, or whether there is a clear hierarchy between them.

\subsection{Future Measurements}
\label{sec:FutureMeasurements}
Using the results of the fit to $\delta$, bounds on all LFV meson and tau decays listed in Table~\ref{tab:expbounds} can now be set. We choose to do this via Monte-Carlo sampling of the likelihood function, and for the specific choice of ${\mu_m = 91.2\text{ GeV}}$ and ${\mu_{\rm low}} = 5\text{ GeV}$. The corresponding limits are summarized in Table~\ref{tab:finalbounds}. 

\begin{table}[h]
    \centering   \renewcommand{\arraystretch}{1.5}
    \begin{tabular}{c|c|l}
    Observable & Prediction & \hfill Current Bound \\ \hline
    $\mathcal{B}(B_s \to \tau\tau)$ & $\big(8.3_{-4.1}^{+12.6}\big)  \times 10^{-5}$ & \qquad $\times 25$ \\
    $\mathcal{B}(B\to K\tau\tau)^{[15,22]}$ & $\big(1.5^{+2.3}_{-0.7}\big)  \times 10^{-5}$ & \, See Sec. \ref{sec:FutureMeasurements} \\
    $\mathcal{B}(B\to K^*\tau\tau)^{[15,19]}$ & $\big(1.3^{+2.0}_{-0.7}\big)  \times 10^{-5}$ & \, See Sec. \ref{sec:FutureMeasurements} \\
    $\mathcal{B}(B_s\to \phi\tau\tau)^{[15,18.8]}$ & $\big(1.2^{+1.9}_{-0.6}\big)  \times 10^{-5}$ & \, See Sec. \ref{sec:FutureMeasurements} \\\hline
    $\mathcal{B}(\tau \to \mu \gamma)$ & $<8.4 \times 10^{-11}$ & \qquad $\times 700$ \\ 
    $\mathcal{B}(\tau \to \mu e e)$ & $<7.6 \times 10^{-10}$ & \qquad $\times 30$ \\
    $\mathcal{B}(\tau \to \mu \mu \mu)$ & $<1.2 \times 10^{-9\hphantom{0}}$ & \qquad $\times 20$ \\
    $\mathcal{B}(\tau \to \mu \rho)$ & $<1.6 \times 10^{-9}$ & \qquad $\times 10$ \\
    $\mathcal{B}(\tau \to \mu \phi)$ & $<9.8 \times 10^{-10}$ & \qquad $\times 30$ \\
    $\mathcal{B}(J/\psi \to \mu \tau)$ & $<3.0 \times 10^{-16}$ & \qquad $\times 10^{10}$ \\
    $\mathcal{B}(\Upsilon \to \mu \tau)$ & $<4.6 \times 10^{-11}$ & \qquad $\times 10^{5}$ \\
    $\mathcal{B}(B_s \to \mu\tau)$  & $<1.8 \times 10^{-7\hphantom{0}}$ & \qquad $\times 200$ \\
    $\mathcal{B}(B \to K \mu\tau)$  & $<2.0 \times 10^{-7\hphantom{0}}$ & \qquad $\times 200$ \\
    $\mathcal{B}(B \to K^* \mu\tau)$  & $<3.3 \times 10^{-7\hphantom{0}}$ & \qquad $\times 70$ \\
    $\mathcal{B}(B_s \to \phi \mu\tau)$  & $<3.6 \times 10^{-7\hphantom{0}}$ & \qquad $\times 40$ \\
    \end{tabular}
    \caption{Bounds on the LFV tau and meson decays most relevant in our flavor setup resulting from our analysis, given at 95\% CL. We also include a prediction for the branching ratio of several ${b\to s(d) \ell \ell}$ channels
    , for which we specify the $q^2$-range as a superscript (in GeV$^2$). In the rightmost column, we quote the approximate relative factor of improvement with respect to the current bounds in Table~\ref{tab:expbounds}.}
    \label{tab:finalbounds}
\end{table}

In order to help visualize the potential of each channel to become the leading probes in constraining the $U(2)_{\ell}$--breaking spurion, we also quote the factor by which our analysis reduces the bounds with respect to the current experimental limits.

Let us remark that if future experimental limits on any of the considered LFV observables fall significantly below our predictions, those observables will become the dominant constraint on the $U(2)_{\ell}$--breaking spurion. For instance, if Belle II reaches the expected $\tau \to \mu \mu \mu$ limit of~\cite{banerjee2022snowmass2021whitepaper,Asadi:2025dii}, 
\begin{equation}
    \mathcal{B}(\tau \to \mu \mu \mu) < 3.5 \times 10^{-10} \quad (\text{expected}),
\end{equation}
about three times lower than our prediction, assuming no increase in the precision of $R_{K^{(*)}}$ nor $\mathcal{B}(B_s\to\mu\mu)$, this channel would become the leading constraint on $\delta$, and would push it to a slightly lower value,
\begin{equation}
    |\delta|_{\rm future} < 0.033 \quad ({\rm at}\; 95\% \; {\rm CL})  \;.
\end{equation}
The expected improvement in ${\tau \to \mu e e}$ would provide a similar bound, albeit slightly higher, while the one in ${\tau \to \mu \gamma}$ would still not meet our predicted limit. 

Other potentially sensitive decays include tau decays into light-generation mesons, exemplified in our case by $\tau \to \mu \rho$ or $\tau \to \mu \phi$. Even though they currently fall behind in terms of constraining power, the improvement in the experimental limits expected by Belle II~\cite{banerjee2022snowmass2021whitepaper} in both of these channels would put them at the forefront of LFV searches, especially in the case of first-generation mesons ($\tau \to \mu \rho$),
\begin{equation}
    \mathcal{B}(\tau \to \mu \rho) < 5.5 \times 10^{-10} \quad (\text{expected}),
\end{equation}
whose projected reach would be comparable to that of $\tau \to \mu \mu \mu$ and $\tau \to \mu e e$. 

In contrast, the current bounds on quarkonium LFV decays, such as ${\Upsilon \to \mu \tau}$ or ${J/\psi \to \mu \tau}$ lie significantly far from the predicted bounds~\cite{banerjee2022snowmass2021whitepaper}:
\begin{align}
    \mathcal{B}(\Upsilon \to \mu \tau) &< 2.7 \times 10^{-6} \quad (\text{expected}),
    \\ 
    \mathcal{B}(J/\psi \to \mu \tau) &< 2.0 \times 10^{-6} \quad (\text{expected}).
\end{align}
We therefore do not expect these observables to become the most constraining ones in the near future, at least as far as our framework is concerned. 

Regarding LFV $B$-meson decays, there are no clear prospects for future experimental sensitivity. This being said, even assuming a one order of magnitude worth of improvement in the bounds, our predictions would still lie mostly out of reach. 
This case, however, is not as drastic as for quarkonium decays: the channel with the closest current bound to our prediction, about 40 times larger, is ${B_s\to\phi\mu\tau}$, and a factor 10 of improvement in the experimental sensitivity would then leave it in the same ballpark as our prediction. Overall, for the moment, we expect LFV $B$-meson decays to remain secondary in terms of their potential to characterize lepton-flavor patterns.

As for the LFV bounds themselves, our predictions tend to lie on the more restrictive side when compared to most EFT-based studies of LFV observables available in the literature~(see e.g.~\cite{Calibbi:2015kma,Husek:2020fru,Descotes-Genon:2023pen,Calibbi_2022,Plakias:2023esq,Fernandez-Martinez:2024bxg}), especially for $B$-meson and quarkonium decays. This occurs because, in our EFT setup, LFV is strongly correlated with the tightly constrained $b\to s\mu\mu$ sector. For many $\tau$-decay channels, however, our resulting upper limits lie instead significantly closer to the least restrictive values found in more dedicated analyses~\cite{Husek:2020fru,Fernandez-Martinez:2024bxg}, mostly due to third-generation penguin-loop effects in the SMEFT-LEFT matching, which soften the impact of these correlations.

We also include the prediction for the branching ratio of ${B_s\to\tau\tau}$, which is mostly independent of $\delta$ and thus fixed by the results of~\cite{Allwicher:2024ncl}, but had not been presented there, given its central role in most frameworks based on NP coupled predominantly to the third generation with an underlying $U(2)^5$ flavor symmetry. The central value we find is about $100$ times greater than the SM prediction~\cite{Bobeth:2013uxa}, and yields an upper limit ($95\%$ CL) about $25$ times below the current experimental bound. 

For completeness, we have also studied the branching ratios of the four-body decays ${B\to K^{(*)}\tau\tau}$ and ${B_s\to \phi\tau\tau}$, choosing a $q^2$-range that avoids the $\psi(2S)$ resonance just as in~\cite{Capdevila:2017iqn}. We find a similar picture, with central values enhanced by two orders of magnitude with respect to the SM expectations~\cite{Capdevila:2017iqn}, and implied upper limits sitting around ${5 \times 10^{-5}}$ ($\geq 95\%$ CL), lying below the current experimental bounds~\cite{ParticleDataGroup:2024cfk,Belle-II:2025lwo} by a factor of about $\times40$ (for both $B\to K\tau\tau$ and $B\to K^*\tau\tau$). However, the recent LHCb indirect bounds in~\cite{LHCb:2025lcw} put much lower limits, corresponding to factors of $\times11$ (for $B\to K^*\tau\tau$) and $\times6$ (for $B_s\to\phi\tau\tau$).
Even though this paints a promising picture for $b \to s \tau\tau$, and much work is being devoted to these channels, significant improvement in the experimental sensitivity to di-tau final states is difficult to achieve and not expected to meet the predictions we provide\footnote{Still, the sensitivities for $b\to s \tau \tau$ decays expected by LHCb at the end of Upgrade II would fall in a similar ballpark as our predictions.} in the near future~\cite{LHCb:2018roe,Belle-II:2022cgf}. Future $Z$-factory experiments, however, such as FCC-ee, have been pointed out as ideal candidates to probe well below the bounds we provide~\cite{Li:2020bvr,Allwicher:2025bub}.

We expect an analogous situation in the rest of ${b\to s(d) \tau\tau}$ channels, including the recently-analyzed ${\Lambda_b \to \Lambda \tau\tau}$~\cite{Bordone:2025elp}.

Concerning $b\to s \mu\mu$ processes, after Upgrade~II and with about $300~\mathrm{fb}^{-1}$ of integrated luminosity collected by LHCb (expected in the early 2040s), the branching fraction $\mathcal{B}(B_s \to \mu \mu)$ is expected to be measured with a relative uncertainty of approximately $5\%$. 
For the LFU ratios $R_{K^{(*)}}$, a precision at the level of $\sim 1\%$ is anticipated~\cite{ATLAS:2025lrr}. With these improvements, the bound on $\delta$ could potentially improve to
    \begin{equation}
    |\delta|_{\rm future} < 0.030 \quad ({\rm at}\; 95\% \; {\rm CL})  \;,
\end{equation}
confirming that these observables are expected to remain among the leading constraints. Regarding the spurion ratio $|\tilde V_\ell|/|\tilde V_q|$, the qualitative discussion would remain mostly unchanged.

\section{Conclusions}
\label{sec:conclusions}

The starting point of this paper was the result obtained in~\cite{Allwicher:2024ncl}. Here the authors showed that a SMEFT framework based on new physics coupled predominantly to the third generation, and an underlying $U(2)^5$ flavor symmetry in the light fermion families broken exclusively in the $U(2)_q$ direction by a spurion term $\tilde V_q$, can describe, with a minimal set of free parameters, current experimental data including all flavor observables exhibiting deviations from the SM predictions. We have extended this framework by including an additional spurion $\tilde V_\ell$ breaking the $U(2)_{\ell}$ approximate flavor symmetry, related to the mixing of left-handed leptons. This addition allowed us to study the bounds imposed on the magnitude of such a breaking by current data on channels mixing third- and second-generation leptons, predominantly $R_{K^{(*)}}$ and ${\mathcal{B}(B_s\to\mu^+\mu^-)}$.

The main result of our analysis is the upper limit on the allowed parameter space of $|\tilde V_\ell|$. This limit, which is discussed in detail in Section~\ref{sec:numanalysis},
turns out to be comparable to the qualitative assessment $$|\tilde V_\ell| \lesssim |V_{ts}|.$$ 

Our analysis provides also an upper bound on the ratio of the sizes of the spurions $\tilde V_\ell$ and $\tilde V_q$, which ensures ${|\tilde V_\ell| < |\tilde V_q|}$, but is not enough to discern between the similar-size scenario, ${|\tilde V_\ell| \sim |\tilde V_q|}$, and a more hierarchical one ${|\tilde V_\ell| \ll |\tilde V_q|}$.
These bounds are obtained under the hypothesis of full alignment of the lepton spurion
to the second generation, and become much 
stronger if $\tilde V_\ell$ has a different 
alignment with non-vanishing component on the first generation. Furthermore, a setup based on an up-aligned basis for the quark doublet in the SMEFT, instead of down-alignment, could also lead to stronger bounds. 

Concerning the dependence on $\Lambda_{\rm NP}$ of our setup, while a different choice would certainly impact the numerical values of $C_{\ell q}^\pm$ resulting from the fit in~\cite{Allwicher:2024ncl}, it would leave the $U(2)_{\ell,q}$-breaking parameters mostly unaffected, barring only RGE effects due to running up to a higher (or lower) $\mu$-scale, expected to be at the few-percent level.

Given the model independence of our setup, these bounds may be used to constrain a wide class of NP models addressing the flavor structure of the SM while also resolving the tensions in the $B$-meson channels. 
These models include, for instance, leptoquark realizations such as the vector $U_1$~\cite{Cornella:2021sby,Allwicher:2024ncl,Crivellin:2025qsq,Calibbi:2015kma}, or the scalar $S_1$ and $S_3$ (e.g.~\cite{Bauer:2015knc,Crivellin:2017zlb,Gherardi:2020det,Crivellin:2025qsq}), extensions of the SM featuring vector-like fermions (e.g.~\cite{Ishiwata:2015cga,Bobeth:2016llm,Alves:2023ufm}), or more elaborate constructs such as composite-Higgs models~\cite{ Barbieri:2012tu, Matsedonskyi:2014iha, Panico:2016ull, Covone:2024elw}. An approximate $U(2)^5$ flavor symmetry and the hypothesis of mainly third-generation NP are easily generated in all these models.

Furthermore, we have translated the bound on $\tilde V_\ell$ into predictions for upper limits on the branching ratios in a representative set of charged-lepton LFV decays, and used them to discuss the potential impact of future improvements in experimental sensitivity. We find that,
if the Belle II expected limits on LFV tau decays~\cite{banerjee2022snowmass2021whitepaper} are met, the four channels of $\tau$-decay into $\mu\mu\mu$, $\mu e e$, $\mu \rho$ and $\mu \phi$ (the latter to a lesser degree) could become leading probes in constraining the breaking of $U(2)_{\ell}$, pushing the bound on $|\tilde V_\ell|$ about $40\%$ lower. These channels would compete on equal grounds with the predicted improvements in $R_{K^{(*)}}$ and ${\mathcal{B}(B_s\to\mu^+\mu^-)}$ at the end of Upgrade II of LHCb~\cite{ATLAS:2025lrr}, which would reduce the bound by a similar amount.

To conclude, we stress that future experimental searches below the bounds reported in Table~\ref{tab:finalbounds} could bring to a positive evidence of LFV in our framework. This would happen if the size of the $U(2)_\ell$--breaking spurion were just below the current bound. This is a well-motivated  possibility in UV models with quark-lepton unification or, more generally, in models where the $U(2)_\ell$ and $U(2)_q$ spurions are of similar size. Conversely, evidences of LFV rates above the bounds in Table~\ref{tab:finalbounds} would disprove this setup.

\subsection*{Acknowledgments}
We would like to thank Gino Isidori for suggesting this project and for many useful discussions. We are grateful to Marzia Bordone for insightful comments on the manuscript. We also thank Lukas Allwicher, Monika Blanke, Xiyuan Gao, Ulrich Nierste, and Peter Stoffer for helpful comments.

S.C. and A.T. acknowledge funding by the Swiss National Science Foundation, projects No.~PCEFP2-194272 and 2000-1-240011. P.M. acknowledges funding by the Spanish MCIN/AEI/\\ 10.13039/501100011033: grant PRE2022-103999 funded by MCIN/AEI/10.13039/501100011033 and by ``ESF Investing in your future", grant CEX2019-000918-M through the “Unit of Excellence Mar\'ia de Maeztu 2020-2023” award to the Institute of Cosmos Sciences.

\appendix
\allowdisplaybreaks
\section{Expressions for the Observables}
\label{app:ObsTh}
In this Appendix, we collect the theoretical expressions for all observables used throughout the paper, written in terms of the WCs of the Lagrangian of the LEFT~\cite{Jenkins:2017dyc}. For that, we focus solely on the Wilson coefficients relevant in our EFT framework, i.e. those receiving non-negligible mixing upon matching and RGE evolution down to the low-energy scale, which corresponds to the list provided in~\ref{app:Wilsons}.

Unless stated otherwise, these expressions are calculated at tree-level in the LEFT. The only observable for which we have deemed convenient to include one-loop effects is the branching ratio for the LFV radiative decay ${\tau \to \mu \gamma}$, given the small size of the dipole Wilson coefficients in the LEFT due to our EFT setup.

\subsection{\texorpdfstring{LFU tests in $b \to s \ell \ell$}{}}\label{lfuapp}
For the expressions for $R_{K^{(*)}}$ in the low-$q^2$, central-$q^2$ and high-$q^2$ (only $R_K$) bins, defined as 
\begin{equation}
     R_{K^{(*)}}{[q^2_{\rm min},q^2_{\rm max}]} = \frac{\int_{q^2_{\rm min}}^{q^2_{\rm max}} \frac{d}{dq^2}\mathcal{B}(B\to K^{(*)} \mu^+\mu^-)}{\int_{q^2_{\rm min}}^{q^2_{\rm max}} \frac{d}{dq^2}\mathcal{B}(B\to K^{(*)} e^+e^-)} \;,
\end{equation}
we have chosen to use \texttt{Flavio}~\cite{Straub:2018kue} to derive pseudo-analytic expressions in terms of the low-energy Wilson coefficients for the ${B\to K^{(*)} \ell^+\ell^-}$ differential branching ratios. We then write their expressions as:
\begin{equation}
    \frac{R_{K^{(*)}}{[q^2_{\rm min},q^2_{\rm max}]}}{R_{K^{(*)}}{[q^2_{\rm min},q^2_{\rm max}]}_{\rm SM}} = \frac{\Delta\mathcal{B}_{2K^{(*)}}^{[q^2_{\rm min},q^2_{\rm max}]}}{\Delta\mathcal{B}_{1K^{(*)}}^{[q^2_{\rm min},q^2_{\rm max}]}} \;,
\end{equation}
where we have defined
\begin{align*}
    \Delta\mathcal{B}_{iK}^{[q^2_{\rm min},q^2_{\rm max}]} \approx \;& \xi_{iK}^{[q^2_{\rm min},q^2_{\rm max}]} + a_{iK}^{[q^2_{\rm min},q^2_{\rm max}]} \, \text{Re}\left\{{L_{\substack{ed\\ii23}}^{V,LL}} \right\}\\
    & + b_{iK}^{[q^2_{\rm min},q^2_{\rm max}]} \, \Big|L_{\substack{ed\\ii23}}^{V,LL}\Big|^2 \\
    & + c_{iK}^{[q^2_{\rm min},q^2_{\rm max}]} \, \Big|L_{\substack{de\\23ii}}^{V,LR}\Big|^2 \;, \\[3mm]
    \Delta\mathcal{B}_{iK^*}^{[q^2_{\rm min},q^2_{\rm max}]} \approx \;& \xi_{i^*K}^{[q^2_{\rm min},q^2_{\rm max}]} + a_{iK^*}^{[q^2_{\rm min},q^2_{\rm max}]} \, \text{Re}\left\{{L_{\substack{ed\\ii23}}^{V,LL}} \right\}\\
    & + b_{iK^*}^{[q^2_{\rm min},q^2_{\rm max}]} \, \Big|L_{\substack{ed\\ii23}}^{V,LL}\Big|^2 \\
    & + d_{iK^*}^{[q^2_{\rm min},q^2_{\rm max}]} \, \text{Re}\left\{{L_{\substack{ed\\ii23}}^{V,LR}} \right\}\\
    & + c_{iK^*}^{[q^2_{\rm min},q^2_{\rm max}]} \, \Big|L_{\substack{de\\23ii}}^{V,LR}\Big|^2 \;.
\end{align*}
As for the numerical coefficients, we first provide the dimensionless constants:
\begin{align*}
    \xi_{2K}^{[0.1,1.1]} = \xi_{1K}^{[0.1,1.1]} = \,& 7.8 \times 10^{-3} \;, \\
    \xi_{2K}^{[1.1,6.0]} = \xi_{1K}^{[1.1,6.0]} = \,& 7.4 \times 10^{-3} \;, \\
    \xi_{2K}^{[14.3,22.9]} = \xi_{1K}^{[14.3,22.9]} = \,& 0.015 \;, \\
    \xi_{2K^*}^{[0.1,1.1]} = \xi_{1K}^{[0.1,1.1]} = \,& 0.786 \;, \\
    \xi_{2K^*}^{[1.1,6.0]} = \xi_{1K}^{[1.1,6.0]} = \,& 0.231 \;.
\end{align*}
The ``$a$'' coefficients read:
\begin{gather*}
    a_{2K}^{[0.1,1.1]} = a_{1K}^{[0.1,1.1]} = 2.69 \times 10^{7} \;, \\
    a_{2K}^{[1.1,6.0]} = a_{1K}^{[1.1,6.0]} = 2.62 \times 10^{7} \;, \\
    a_{2K}^{[14.3,22.9]} = a_{1K}^{[14.3,22.9]} = 2.34 \times 10^{7} \;, \\
    a_{2K^*}^{[0.1,1.1]} = 3.92 \times 10^{7} \;, \quad a_{1K^*}^{[0.1,1.1]} = 3.89 \times 10^{7} \;, \\
    a_{2K^*}^{[1.1,6.0]} = 8.38 \times 10^{7} \;, \quad a_{1K^*}^{[1.1,6.0]} = 8.36 \times 10^{7} \;.
\end{gather*}
The ``$b$'' coefficients:
\begin{gather*}
    b_{2K}^{[0.1,1.1]} = b_{1K}^{[0.1,1.1]} = 2.38 \times 10^{16} \;, \\
    b_{2K}^{[1.1,6.0]} = b_{1K}^{[1.1,6.0]} = 2.38 \times 10^{16} \;, \\
    b_{2K}^{[14.3,22.9]} = b_{1K}^{[14.3,22.9]} = 2.32 \times 10^{16} \;, \\
    b_{2K^*}^{[0.1,1.1]} = 9.73 \times 10^{15} \;, \quad b_{1K^*}^{[0.1,1.1]} = 9.67 \times 10^{15} \;, \\
    b_{2K^*}^{[1.1,6.0]} = b_{1K^*}^{[1.1,6.0]} = 2.71 \times 10^{16} \;.
\end{gather*}
The ``$c$'' coefficients read:
\begin{gather*}
    c_{2K}^{[0.1,1.1]} = c_{1K}^{[0.1,1.1]} = 2.39 \times 10^{16} \;, \\
    c_{2K}^{[1.1,6.0]} = c_{1K}^{[1.1,6.0]} = 2.38 \times 10^{16} \;, \\
    c_{2K}^{[14.3,22.9]} = c_{1K}^{[14.3,22.9]} = 2.32 \times 10^{16} \;, \\
    c_{2K^*}^{[0.1,1.1]} = 9.74 \times 10^{15} \;, \quad c_{1K^*}^{[0.1,1.1]} = 9.68 \times 10^{15} \;, \\
    c_{2K^*}^{[1.1,6.0]} = c_{1K^*}^{[1.1,6.0]} = 2.71 \times 10^{16} \;.
\end{gather*}
Finally the ``$d$'' coefficients read:
\begin{gather*}
    d_{2K^*}^{[0.1,1.1]} = 2.78 \times 10^{7} \;, \quad d_{1K^*}^{[0.1,1.1]} = 2.75 \times 10^{7} \;, \\
    d_{2K^*}^{[1.1,6.0]} = 5.20 \times 10^{7} \;, \quad d_{1K^*}^{[1.1,6.0]} = 5.18 \times 10^{7} \;.
\end{gather*}

\subsection{\texorpdfstring{$b \to s \ell \ell$ decays}{}}\label{bsllapp}
We include the branching ratios for the decays ${B_s \to \mu \mu}$ and ${B_s \to \tau \tau}$,
\begin{align}
    \frac{\mathcal{B}(B_s \to \mu \mu)}{\mathcal{B}(B_s \to \mu \mu)_{\rm SM}} = \;&  \xi_\mu \, \Big|L_{\substack{ed\\2223}}^{V,LL} - L_{\substack{de\\2322}}^{V,LR}\Big|^2  \;,
    \\
    \mathcal{B}(B_s \to \tau \tau) = \;& a_\tau \, \Big|L_{\substack{ed\\3323}}^{V,LL} - L_{\substack{de\\2333}}^{V,LR}\Big|^2 \;,
\end{align}
with coefficients
\begin{equation*}
      \xi_\mu = 1.86 \times 10^{16} \text{ GeV}^2 \;, \quad a_\tau = 1.55 \times 10^{10}\text{ GeV}^2 \;.
\end{equation*}

We also consider the branching ratios of ${B \to K^{(*)} \tau \tau}$ and ${B_s \to \phi \tau \tau}$ integrated over the $q^2$ bins indicated in the square brackets (in GeV$^2$), chosen to avoid the $\psi(2S)$ resonance,
\begin{align}
    \mathcal{B}(B \to K \tau \tau)&^{[15,22]} = \xi_{\tau K} + a_{\tau K} \, \text{Re}\left\{{L_{\substack{ed\\3323}}^{V,LL}}\right\} \\
    &\hspace{-5mm}+ b_{\tau K} \, \text{Im}\left\{{L_{\substack{ed\\3323}}^{V,LL}}\right\} + \tilde c_{\tau K} \Big|L_{\substack{ed\\3323}}^{V,LL}\Big|^2 \;, \nonumber\\
    \mathcal{B}(B \to K^* \tau \tau)&^{[15,19]} = \xi_{\tau K^*} + a_{\tau K^*} \, \text{Re}\left\{{L_{\substack{ed\\3323}}^{V,LL}} \right\}\\
    &\hspace{-5mm}+ b_{\tau K^*}\, \text{Im}\left\{{L_{\substack{ed\\3323}}^{V,LL}}\right\} + \tilde c_{\tau K^*} \Big|L_{\substack{ed\\3323}}^{V,LL}\Big|^2 \;, \nonumber\\
    \mathcal{B}(B \to \phi \tau \tau)&^{[15,18.8]} = \xi_{\tau \phi} + a_{\tau \phi} \, \text{Re}\left\{{L_{\substack{ed\\3323}}^{V,LL}}\right\} \\
    &\hspace{-5mm}+ b_{\tau \phi} \, \text{Im}\left\{{L_{\substack{ed\\3323}}^{V,LL}}\right\} + \tilde c_{\tau \phi} \Big|L_{\substack{ed\\3323}}^{V,LL}\Big|^2 \;, \nonumber
\end{align}
and the coefficients read:
\begin{align*}
      \xi_{\tau K} =\;& 1.59 \times 10^{-9} \;, \quad & a_{\tau K} =\;& 3.22 \text{ GeV} \;, \\
      b_{\tau K} =\;& 0.05 \text{ GeV} \;, \quad & c_{\tau K} =\;& 2.79 \times 10^{9} \text{ GeV}^2 \;,\\
      \xi_{\tau K^*} =\;& 3.08 \times 10^{-9} \;, \quad & a_{\tau K^*} =\;& 4.20 \text{ GeV} \;, \\
      b_{\tau K^*} =\;& 0.07 \text{ GeV} \;, \quad & c_{\tau K^*} =\;& 2.57 \times 10^{9} \text{ GeV}^2 \;,\\      
      \xi_{\tau \phi} =\;& 2.62 \times 10^{-9} \;, \quad & a_{\tau \phi} =\;& 3.60 \text{ GeV} \;, \\
      b_{\tau \phi} =\;& 0.06 \text{ GeV} \;, \quad & c_{\tau \phi} =\;& 2.37 \times 10^{9} \text{ GeV}^2 \;,
\end{align*}
and we have neglected the (very suppressed) contributions from $L_{\substack{de\\2333}}^{V,LR}$.

\subsection{\texorpdfstring{LFV $B$-meson decays}{}}
We list the expressions for several $B$-meson decays featuring lepton-flavor violating final states, namely ${B_s \to \mu \tau}$, ${B_s \to K^{(*)} \mu \tau}$, ${B_s \to \phi \mu \tau}$. In all cases, we include both possibilities for the lepton pair, i.e. $\mu^-\tau^+$ and $\mu^+\tau^-$. Their expressions are especially simple in our EFT framework, given that they are mediated by a single relevant Wilson coefficient, yielding:
\begin{align}
    \mathcal{B}(B_s \to \mu\tau) = a_{B_s} \, \bigg(\Big|L_{\substack{ed\\2323}}^{V,LL}\Big|^2 + \Big|L_{\substack{ed\\3223}}^{V,LL}\Big|^2\bigg) \;, \\
    \quad \mathcal{B}(B_s \to \phi\mu\tau) = b_{B_s} \, \bigg(\Big|L_{\substack{ed\\2323}}^{V,LL}\Big|^2 + \Big|L_{\substack{ed\\3223}}^{V,LL}\Big|^2\bigg) \;, \\
    \quad \mathcal{B}(B \to K\mu\tau) = a_{K} \, \bigg(\Big|L_{\substack{ed\\2323}}^{V,LL}\Big|^2 + \Big|L_{\substack{ed\\3223}}^{V,LL}\Big|^2\bigg) \;,\\
    \quad \mathcal{B}(B \to K^*\mu\tau) = a_{K^*} \, \bigg(\Big|L_{\substack{ed\\2323}}^{V,LL}\Big|^2 + \Big|L_{\substack{ed\\3223}}^{V,LL}\Big|^2\bigg) \;,
\end{align}
with coefficients
\begin{gather*}
      a_{B_s} = 8.22 \times 10^{9} \text{ GeV}^2 \;, \quad b_{B_s} = 1.59 \times 10^{10} \text{ GeV}^2 \;, \\
      a_{K} = 8.67 \times 10^{9} \text{ GeV}^2 \;, \quad a_{K^*} = 1.55 \times 10^{10} \text{ GeV}^2 \;.
\end{gather*}
We have derived the expression for the branching ratio of $B_s \to \phi \mu \tau$ from \texttt{Flavio}~\cite{Straub:2018kue}.

\subsection{LFV quarkonium decays}
We consider the branching ratios of the $\Upsilon(1S)$ and $J/\psi$ quarkonium states into $\mu \tau$, for which we combine both $\mu^-\tau^+$ and $\mu^+\tau^-$. Their expressions read:
\begin{align}
    \mathcal{B}(\Upsilon \to \mu \tau) = \;& a_\Upsilon \, \Big|L_{\substack{ed\\2333}}^{V,LL} + L_{\substack{ed\\2333}}^{V,LR} \Big|^2 \;,
    \\
    \mathcal{B}(J/\psi \to \mu \tau) = \;& a_{J/\psi} \, \Big|L_{\substack{eu\\2322}}^{V,LL} + L_{\substack{eu\\2322}}^{V,LR} \Big|^2 \;,
\end{align}
with coefficients
\begin{equation*}
      a_\Upsilon = 4.15 \times 10^4 \text{ GeV}^2 \;, \quad a_{J/\psi} = 194.03 \text{ GeV}^2 \;.
\end{equation*}
We also include here LFV $\tau$ decays \textit{into} quarkonia, namely $\tau \to \mu \rho$ and $\tau \to \mu \phi$. Their expressions read:
\begin{align}
    \mathcal{B}(\tau \to \mu \phi) = \;& a_\phi \, \Big|L_{\substack{ed\\2322}}^{V,LL} + L_{\substack{ed\\2322}}^{V,LR} \Big|^2 \;,
    \\
    \mathcal{B}(\tau \to \mu \rho) = \;& a_\rho \, \Big|L_{\substack{ed\\2311}}^{V,LL} + L_{\substack{ed\\2311}}^{V,LR} - L_{\substack{eu\\2311}}^{V,LL} - L_{\substack{eu\\2311}}^{V,LR} \Big|^2 \;,
\end{align}
with coefficients
\begin{equation*}
      a_\phi = 2.63 \times 10^8 \text{ GeV}^2 \;, \quad a_{\rho} = 1.21 \times 10^8 \text{ GeV}^2 \;.
\end{equation*}
For all these expressions, we have neglected the (essentially vanishing) imaginary part in the Wilson coefficients; i.e. for $k=1,2,3$,
\begin{equation*}
    L_{\substack{ed\\23kk}}^{V,LL} = \Big(L_{\substack{ed\\32kk}}^{V,LL}\Big)^* \;, \qquad 
    L_{\substack{ed\\23kk}}^{V,LR} = \Big(L_{\substack{ed\\32kk}}^{V,LR}\Big)^* \;,
\end{equation*}
and thus the combination of the $\mu^-\tau^+$ and $\mu^+\tau^-$ final states ends up adding up to a factor of 2.

\subsection{\texorpdfstring{LFV $\tau$ decays}{}}
We include here the branching ratios of different non-hadronic LFV $\tau$ decays, including radiative decays (${\tau \to \mu \gamma}$) and fully leptonic decays (${\tau \to \mu \mu \mu}$ and ${\tau \to \mu e e}$). For the radiative decays, we include both the tree-level amplitudes with dipole operators and the one-loop penguin-diagram contributions from the four-lepton operators. The final expressions read:
\begin{align}
    \mathcal{B}(\tau \to \mu \gamma) = \; & a_\gamma \, \bigg( \, \Big|L_{\substack{\vphantom{d}e\gamma\\23}} + \frac{e\,m_\tau}{16\pi^2}L_{\substack{ee\\2333}}^{V,LR} \Big|^2 \nonumber \\[2mm]
    & \hspace{1cm} + \Big|L_{\substack{\vphantom{d}e\gamma\\32}}^{*} + \frac{e\,m_\mu}{16\pi^2}L_{\substack{ee\\2322}}^{V,LR} \Big|^2 \, \bigg) \;,
    \\
    \mathcal{B}(\tau \to \mu \mu \mu) = \;& a_{\mu} \, \Big|L_{\substack{ee\\2223}}^{V,LL}\Big|^2 + b_{\mu} \, \Big|L_{\substack{ee\\2322}}^{V,LR}\Big|^2 \nonumber \\
    & + c_{\mu} \, \text{Re}\left\{{\Big(L_{\substack{ee\\2223}}^{V,LL}\Big)^* L_{\substack{ee\\2322}}^{V,LR}}\right\} \;,
    \\
    \mathcal{B}(\tau \to \mu e e) = \;& a_e \, \Big|L_{\substack{ee\\1123}}^{V,LL}\Big|^2 + b_e \, \Big|L_{\substack{ee\\2311}}^{V,LR}\Big|^2 \;,
\end{align}
with coefficients
\begin{align*}
      a_\gamma = \,& 1.95 \times 10^{11} \text{ GeV} \;, \quad & a_{3\mu} = \,& 1.21 \times 10^9 \text{ GeV}^2 \;, \\
       \quad b_{3\mu} = \,& 1.59 \times 10^8 \text{ GeV}^2 \;, \quad & c_{3\mu} = \,& 1.64 \times 10^7 \text{ GeV}^2 \;,\\
       a_{e} = \,& 2.48 \times 10^8 \text{ GeV}^2 \;, \quad & b_{e} = \,& 1.55 \times 10^8 \text{ GeV}^2 \;.
\end{align*}

\section{Low-Energy Wilson Coefficients}
\label{app:Wilsons}
We list here the leading expressions (neglecting terms or order $2\%$ or smaller) of all Wilson coefficients in the LEFT relevant for our study, evaluated at $\mu = 5$ GeV, as obtained from \texttt{DsixTools}~\cite{Fuentes-Martin:2020zaz}, in terms of the free parameters in our UV setup. For these expressions, we use a matching scale of $\mu = 91.2$ GeV, and substitute all CKM matrix elements for their values according to the Wolfenstein parameterization and the parameters in Table~\ref{tab:SMvalues}. All coefficients that might contribute to the different observables, but we do not cite, should be understood as much smaller than the level of precision chosen for these expressions.

As a general rule to schematically understand the origin of each type of contribution in these expressions, one must take into account that terms scaling as expected of Eq.~(\ref{eq:SMEFTscaling}) and the fermion flavors involved will be dominated by the tree-level matching to the SMEFT (even though they can include non-negligible one-loop corrections), while those violating the naive scaling will be stemming from contributions starting at one-loop, both in the matching or in the running.

Starting with the Wilson coefficients relevant for FCNC ${b \to s \ell \ell}$ transitions we have, normalized to~$10^{-9}$~GeV$^{-2}$,
\begin{align}
\label{eq:LedVLL[3323]}
L_{\substack{ed\\3323}}^{V,LL} \approx \; & -7.04 + 0.13 i + (78 - 1.57i) \, C_{\ell q}^+ \,\epsilon\nonumber \\
&  + 1.81 \, C_{\ell q}^+ - 0.45 \, C_{\ell q}^- \;, \\ 
L_{\substack{de\\2333}}^{V,LR} \approx \; & 0.29 - 0.34 \, C_{\ell q}^+ \,\epsilon \;, \\ 
L_{\substack{ed\\2223}}^{V,LL} \approx \; & -7.04 + 0.13 i - (0.33 - 78 \, c_e^2 \, \delta^2) \, C_{\ell q}^+ \,\epsilon \;, \\ 
L_{\substack{de\\2322}}^{V,LR} \approx \; & 0.29 - 0.34 \, C_{\ell q}^+ \,\epsilon \;, \\ 
L_{\substack{ed\\1123}}^{V,LL} \approx \; & -7.04 + 0.13 i - (0.33 - 78 \, s_e^2 \, \delta^2) \, C_{\ell q}^+ \,\epsilon \;, \\ 
\label{eq:LdeVLR[2311]}
L_{\substack{de\\2311}}^{V,LR} \approx \; & 0.29 - 0.34 \, C_{\ell q}^+ \,\epsilon \;.
\end{align}
One can see from Eq.~(\ref{eq:LedVLL[3323]}) that the Wilson coefficient involving left-handed third-family lepton currents gets a direct NP contribution free of $\delta$, dominated by the tree-level matching to the SMEFT, which then mixes down into the lighter families via penguin-loop effects in the RGE and matching (independent of $\delta$) and via the direct mixing due to the spurion $V_{\ell}$ (proportional to $\delta^2$). This contribution also mixes into the coefficients involving right-handed leptons, again due to penguin-loop effects, in a universal way across lepton flavors.

Moving on to the coefficients mediating ${b \to s \tau \mu}$ transitions we find, normalized again to $10^{-9}$~GeV$^{-2}$,
\begin{align}
L_{\substack{ed\\2323}}^{V,LL} \approx \; & 1.82 \, c_e \, \delta \, C_{\ell q}^+ + (78.33 - 1.57 i) \, c_e \, \delta \, C_{\ell q}^+ \epsilon \;.
\end{align}
This case features only one relevant coefficient, receiving two main contributions from the matching to the SMEFT: one dominated by the tree-level matching, scaling as expected of Eq.~(\ref{eq:SMEFTscaling}), and another one appearing at one-loop and scaling as an operator featuring a $\bar t t$ pair.

Next we focus on the Wilson coefficients relevant for ${\bar q q \to \tau \mu}$ or ${\tau \to \mu \bar q q}$ transitions. Normalizing them to $10^{-8}$~GeV$^{-2}$, we find
\begin{align}
\label{eq:LedVLL[2333]}
L_{\substack{ed\\2333}}^{V,LL} \approx \; & (180 C_{\ell q}^+ + 9.13 C_{\ell q}^-)\,c_e\,\delta \;, \\ 
L_{\substack{ed\\2333}}^{V,LR} \approx \; & (1.90 C_{\ell q}^+ - 1.29 C_{\ell q}^-)\,c_e\,\delta \;, \\ 
L_{\substack{ed\\2322}}^{V,LL} \approx \; & \big(\{0.32\,\epsilon^2 -11.74\} C_{\ell q}^+ + 8.03 C_{\ell q}^-\big)\,c_e\,\delta \;, \\
L_{\substack{ed\\2322}}^{V,LR} \approx \; & (1.90 C_{\ell q}^+ - 1.29 C_{\ell q}^-)\,c_e\,\delta \;, \\ 
L_{\substack{ed\\2311}}^{V,LL} \approx \; & (-11.74 C_{\ell q}^+ + 8.03 C_{\ell q}^-)\,c_e\,\delta \;, \\ 
L_{\substack{ed\\2311}}^{V,LR} \approx \; & (1.90 C_{\ell q}^+ - 1.29 C_{\ell q}^-)\,c_e\,\delta \;, \\
L_{\substack{eu\\2322}}^{V,LL} \approx \; & \big(\{10.27 - 0.02\,\epsilon -0.08\,\epsilon^2\} C_{\ell q}^+ \nonumber \\
& \hphantom{\big(} - \{6.7 - 0.81\,\epsilon - 0.37\,\epsilon^2\} C_{\ell q}^- \,\big)\,c_e\,\delta \;, \\ 
L_{\substack{eu\\2322}}^{V,LR} \approx \; & (-3.71 C_{\ell q}^+ + 2.53 C_{\ell q}^-)\,c_e\,\delta \;, \\ 
L_{\substack{eu\\2311}}^{V,LL} \approx \; & (-10.29 C_{\ell q}^+ - 7.11 C_{\ell q}^-)\,c_e\,\delta \;, \\ 
L_{\substack{eu\\2311}}^{V,LR} \approx \; & (-3.71 C_{\ell q}^+ + 2.53 C_{\ell q}^-)\,c_e\,\delta \;.
\end{align}
The picture in this case is somewhat analogous to the situation in Eqs.~(\ref{eq:LedVLL[3323]}-\ref{eq:LdeVLR[2311]}), with third generation left-handed quark currents receiving the largest contribution, and mixing into the coefficients for lighter quarks via an interplay of penguin-loop effects in the matching and running (independent of $\epsilon$), and direct contributions proportional to $\epsilon^2$. This being said, the penguin-loop effects are much more prominent in this case, and thus most coefficients are dominated by the scaling of Eq.~(\ref{eq:LedVLL[2333]}).

Finally, we focus on the Wilson coefficients mediating non-hadronic LFV tau decays. Normalizing them to $10^{-8}$~GeV$^{-2}$ once again, we find
\begin{align}
\label{eq:LeeVLL[2333]}
L_{\substack{ee\\2333}}^{V,LL} \approx \; & (-3.96 C_{\ell q}^+ + 2.72 C_{\ell q}^-)\,c_e\,\delta \;, \\ 
L_{\substack{ee\\2333}}^{V,LR} \approx \; & (5.77 C_{\ell q}^+ - 3.92 C_{\ell q}^-)\,c_e\,\delta \;, \\ 
\label{eq:LeeVLL[2223]}
L_{\substack{ee\\2223}}^{V,LL} \approx \; & (-3.96 C_{\ell q}^+ + 2.72 C_{\ell q}^-)\,c_e\,\delta \;, \\
L_{\substack{ee\\2322}}^{V,LR} \approx \; & (5.77 C_{\ell q}^+ - 3.92 C_{\ell q}^-)\,c_e\,\delta \;, \\ 
\label{eq:LeeVLL[1123]}
L_{\substack{ee\\1123}}^{V,LL} \approx \; & (-1.98 C_{\ell q}^+ + 1.36 C_{\ell q}^-)\,c_e\,\delta \;, \\
L_{\substack{ee\\2311}}^{V,LR} \approx \; & (5.77 C_{\ell q}^+ - 3.92 C_{\ell q}^-)\,c_e\,\delta \;, \\ 
L_{\substack{\vphantom{d}e\gamma\\23}} \approx \; & (5.7 C_{\ell q}^+ - 9.5 C_{\ell q}^-)\,c_e\,\delta \times 10^{-2} \;, \\ 
L_{\substack{\vphantom{d}e\gamma\\32}} \approx \; & (3.4 C_{\ell q}^+ - 5.6 C_{\ell q}^-)\,c_e\,\delta \times 10^{-3} \;.
\end{align}
All of these coefficients are produced starting at one-loop order in the matching to the SMEFT, dominated by penguin-loop effects. As such, their expressions are mostly flavor-universal. The apparent exception of Eq.~(\ref{eq:LeeVLL[1123]}) is in fact a relative factor of 2 due to the symmetries in the flavor indices of Eqs.~(\ref{eq:LeeVLL[2333]}-\ref{eq:LeeVLL[2223]}).

\bibliographystyle{elsarticle-num} 
\bibliography{references}






\end{document}